\documentclass{jetpl}
\twocolumn

\lat

\title{Towards azimuthal anisotropy of direct photons}

\rtitle{Towards azimuthal anisotropy of direct photons\ldots}

\sodtitle{Towards azimuthal anisotropy of direct photons }

\author{V.\,V.\,Goloviznin$^{+}$,
A.\,M.\,Snigirev$^{*}$\/\thanks{e-mail: snigirev@lav01.sinp.msu.ru},
G.\,M.\,Zinovjev$^+$}

\rauthor{V.\,V.\,Goloviznin V.V., A.\,M.\,Snigirev A.M., G.\,M.\,Zinovjev G.M.}

\sodauthor{Goloviznin, Snigirev, Zinovjev}

\address{$^+$ Bogolyubov Institute for Theoretical Physics, National Academy of Sciences of Ukraine, Kiev 03680, Ukraine \\~\\
$^*$ Skobeltsyn Institute of Nuclear Physics, Lomonosov Moscow 
State University, 119991, Moscow, Russia}

\abstract{Intensive radiation of magnetic bremsstrahlung type (synchrotron radiation) resulting from the interaction of escaping quarks with the collective confining colour field is discussed as a new possible mechanism of observed direct photon anisotropy }


\begin{document}

\maketitle
The mighty wealth of experimental data on relativistic heavy ion collisions collected in the different
experiments in recent years (even before putting LHC in operation) is reasonably well described (but less
well understood) in the framework of approach based on the relativistic hydrodynamic equations~\cite{QM2012,heinz}.
In particular, a (nearly) perfect hydrodynamics has successfully predicted an existence of radial and 
elliptic flows, their dependence on centrality, mass, beam energy and transverse momentum. Crucial moment
of this approach is that the respective liquid possesses rather special transport properties. Indeed,
the ratio of its shear viscosity coefficient $\eta$ to the entropy density $s$, i.e. $\eta/s$, develops 
very small magnitude. Obviously, any microscopic interpretation of new experimental data at this energy 
scale should take into account this novel theoretical background but also to answer the most exciting 
question what is that fluid entity.
         
Measuring the photon radiation in ultrarelativistic collisions of heavy nuclei has been suggested as one 
of the most indicative signals of producing new state of matter many years ago~\cite{feinberg,shuryak}.
In this context the recent measurements by the PHENIX Collaboration which show the azimuthal anisotropy 
of produced direct photons very close to the hadron one~\cite{phenix2011} are rather exciting. This result 
appears to be in a serious contradiction with expected dominance of photon production from quark gluon plasma 
at an early stage of ion collision at the top RHIC (Brookhaven) and now available LHC (CERN) energies. The 
observed temperature of ``anomalous'' photon radiation (about $T_{\rm ave}\simeq220$ Mev) is in accordance with 
the PHENIX Collaboration measurements~\cite{phenix2010} at the energy $\sqrt{s}=200$ GeV of heavy ion collisions. 
This temperature magnitude being considered as a result of averaging over the entire evolution of the matter 
created in nuclear collisions is noticeably higher than the phase transition temperature (this statement is 
wandering over the all phenomenological papers albeit we understand the lattice QCD declares the presence 
of a cross-over only~\cite{aoki}) and obviously supports the scenario of photon radiation from quark gluon 
plasma. Forming a gluon condensate which radiates the photons at the early stage of collisions is 
considered~\cite{chiu} as another alternative explanation of high photon source temperature measured.

However, in both these scenarios the photon azimuthal anisotropy is declared to be small~\cite{chatterjee} 
and insufficient to explain the experimental data mentioned. For the time being this new result of the 
PHENIX Collaboration promoted great interest in both experimental and theoretical studies and several 
phenomenological suggestions~\cite{bzdak,hees,kharzeev,liu,linnyk} are under discussion to understand an origin of 
this exciting observation. The main goal of our present letter is to draw attention to another significant 
mechanism that contributes to the observed anisotropy of direct photons and is apparently not taken into 
consideration in the existing theoretical estimates. The reference is to a ``magnetic bremsstrahlung-like 
radiation'' (or synchrotron radiation in present terminology) of quarks in the collective colour field ensuring 
confinement. As it has been argued in our old papers~\cite{gol1,gol2,gol3} such a radiation from the surface 
layer of quark-gluon system produced in collision is intensive enough and comparable with 
volume~\cite{feinberg,shuryak} photon radiation (``Compton scattering of gluons'', $gq \rightarrow \gamma q$ and 
annihilation of quark-antiquark pairs, $q{\bar q} \rightarrow \gamma g$). Quantitatively, an effect is rooted 
in the large magnitude of quark confining force $\sigma \simeq 0.2$~Gev$^2$. 

Theoretically, the basic conditions to have such a radiation available are easily realized as 1) the 
presence of relativistic light quarks ($u$ and $d$ quarks) in the quark gluon system; 2) the semiclassical 
nature of their motion; 3) confinement. Then as a result, each quark (antiquark) at the boundary of the system 
volume moves along a curve trajectory and (as any classical charge undergoes an acceleration) emits photons. 
This radiation which is usually classified as a ``magnetic bremsstraglung'' (synchrotron radiation) will be 
nonisotropic for the noncentral collisions because the photons are dominantly emitted around the direction 
determined by surface normal. Estimating the magnitude of this effect we have utilized~\cite{gol1,gol2,gol3} 
the chromoelectric flux tube model ~\cite{tube1,tube2,tube3} in which the interaction between the volume of 
quark-gluon system and colour object crossing over its boundary develops the constant force $\sigma$ bringing 
a colour object back. Apparently, this force is acting along the normal to the plasma surface. 

A large value of $\sigma$ results in the large magnitude of characteristic parameter
$\chi = ((3/2) \sigma E/m^3)^{1/3}$
(where $E$ and $m$ are the energy and mass of the emitting particle, respectively) for $u$ and $d$ quarks 
(the strong-field case) and then, as Ref~\cite{book} teaches, the radiation intensity becomes independent of 
the particle mass and looks like
\begin{equation}
\label{c1}
dI/dt = 0.37 e_q^2 \alpha(E\sigma \sin \varphi)^{2/3}.
\end{equation}
Here $e_q$ is the quark charge in the units of electron charge, $\alpha$ is the fine structure constant and 
$\varphi$ is the angle between the particle velocity and the normal to the quark-gluon system surface. The 
spectral radiation density may be approximated by
\begin{equation}
\label{c2}
\frac{dI}{d\omega dt} = 0.52 e_q^2 \alpha \omega^{1/3} (\sigma \sin
\varphi/E)^{2/3},~~0\leq \omega < E
\end{equation}
that is quite robust to do the estimates excepting the frequency interval close to $E$. As for the angular 
distribution the ultrarelativistic particles are radiating photons mainly into small ($\sim m/E$) angles 
around the instantaneous direction of their velocity, and the velocity distribution of quarks in the 
quantitative estimates is usually treated as isotropic inside the quark-gluon volume. 

In the approach to the relativistic heavy ion collisions dealing with hydrodynamical scaling solution
~\cite{bjorken} one has a cylindrically symmetric plasma volume that is expanding in the longitudinal 
direction at central collisions. Adapting an ideal gas equation of state for quark-gluon system we have
\begin{equation}
\label{c8}
T = T_0(\tau_0/\tau)^{1/3},
\end{equation}
where $T_0$ is the temperature at the proper time $\tau_0$ of hydrodynamic stage. Thus, the total number of 
photons radiated from the plasma surface at its hydrodynamic evolution can be estimated~\cite{gol2,gol3} 
explicitly as
\begin{eqnarray}
\label{c9}
N_{\rm surface}^{\gamma}& = \int \frac{dN_{\gamma}}{dSdt} 2\pi rdzdt = 2\pi r \int
\frac{dN_{\gamma}}{dSdt}\tau d \tau dy \nonumber \\
&  = A \langle e_q^2 \rangle \alpha \cdot 2Y \cdot 3(\tau_0 T_0^3)^2 \frac{\pi
r^2}{2T_c^2} \frac{4}{7/3 rT_c^{1/3}\sigma^{1/3}}\nonumber\\
&  \times \Bigg[1-\Bigg(\frac{T_c}{T_0}\Bigg)^{7/3}\Bigg], 
\end{eqnarray}
where $T_c$ is the phase-transition temperature, $r$ is the cylinder radius, $2Y$ is the corresponding 
rapidity interval, $\langle e_q^2 \rangle = e_u^2 + e_d^2$, $e_u$ and  $e_d$ are the $u-$ and 
$d-$quark charges, $A = 3.12g\cdot 2^{5/3} \Gamma^2 (4/3) /(2\pi)^2 \simeq 1.2$, $\Gamma$ is the gamma 
function, $g$-spin$\times$colour=6 is the number of quark degrees of freedom.

Evaluating the number of photons coming from the channels $gq \to \gamma q$, $q\bar q \to \gamma g$ 
(from plasma volume) we have~\cite{gol2,gol3}
\begin{eqnarray}
\label{c10}
N_{\rm volume}^{\gamma}& &= \int \frac{dN_{\gamma}}{d^3xdt} \pi
r^2dzdt \\
& & = B \alpha\cdot 2Y\cdot 3(\tau_0T_0^3)^2 \frac{\pi
r^2}{2T_c^2}\Bigg[1-\Bigg(\frac{T_c}{T_0}\Bigg)^2\Bigg].\nonumber 
\end{eqnarray}
where
$B \simeq\frac{5}{144}\pi \alpha_s \ln\frac{1}{\alpha_s}$ as in~\cite{sinha},
$\alpha_s$ is the running coupling constant.

Comparing (\ref{c9}) and (\ref{c10}) we find that the difference between two mechanisms is mainly 
determined by the coefficient $(rT_c^{1/3}\sigma^{1/3})^{-1}$. Taking into account the ratio of constant
quantities $A$ and $B$ we find $N^{\gamma}_{\rm surface}/N_{\rm volume}^{\gamma} \approx 2$ at $r=10$ fm 
and reasonably large initial temperature $T_0$. Thus, we may conclude the intensity of surface radiation 
for the quark-gluon systems of the transverse size $1-10$ fm which are expected to occur in relativistic 
heavy ion collisions is comparable (even larger, especially for noncentral collisions of small transverse
size) with intensity of the volume mechanism of photon production (which is the basic radiation source in
the current theoretical appraisals) even if we deal with $T_c$ around $150$ MeV that corresponds to the 
present-day lattice QCD results~\cite{fodor}. The similar estimation can be obtained~\cite{gol2,gol3} for hard enough photons also.

Obviously, the photon emanation from the surface mechanism of noncentral ion collisions is nonisotropic. 
Indeed, photons are emitted mainly around the direction determined by the normal to the ellipsoid-like 
surface. In the transverse ($x$-$y$) plane (the beam is running along ($z$)-axis) the direction of this 
normal (emitted photons) is determined by the spatial azimuthal angle $\phi_s=\tan^{-1}(y/x)$ as
\begin{equation}
\label{phi}
\tan(\phi_{\gamma}) = (R_x/R_y)^2 \tan(\phi_s).
\end{equation}
The shape of quark-gluon system surface in transverse plane is controlled by the radii 
$R_x= R \sqrt{1-\epsilon}$ and $R_y= R \sqrt{1+\epsilon}$ with the eccentricity $\epsilon =b/2R_A$ 
($b$ is the impact parameter, $R_A$ is the radius of the colliding (identical) nuclei).
The photon azimuthal anisotropy can be characterized by the second Fourier component
\begin{equation}
\label{v2}
v_2^{\gamma} = \frac{\int d\phi_{\gamma}\cos(2\phi_{\gamma})(dN^{\gamma}/d\phi_{\gamma})}{\int d\phi_{\gamma}(dN^{\gamma}/d\phi_{\gamma})}
\end{equation}
and is proportional to the ``mean normal''
\begin{equation}
\label{v2cos}
v_2^{\gamma} \propto\frac{\int d\phi_s \cos(2\phi_{\gamma})} {2\pi}= \epsilon.
\end{equation}

Summarizing we would like to maintain positively that the surface mechanism of photon production is 
intensive enough, develops the azimuthal anisotropy and is capable of resolving the PHENIX direct photons 
puzzle~\cite{phenix2011} still without appealing to the non-equilibrium dynamics of heavy ion collision process
and quantitatively is enough flexible to absorb the news of ``changing landscape'' of lattice QCD calculations.

We are very thankful to Kyrill Bugaev, Igor Dremin, Paolo Giubellino, Serguei Molodtsov and Bikash Sinha for encouraging discussions.
Friendly criticism by Zoltan Fodor is greatly appreciated, it helped to improve the presentation, indeed. 
This work is partly supported 
by Russian Foundation for Basic Research Grants No. 13-02-01005 
and 
the Grant of Fundamental Physics Programme launched by Division of Physics and Astronomy (National Academy of 
Sciences of Ukraine).


%

\end{document}